\documentstyle[12pt]{article}
\makeatother
\begin{document}

\begin{center}
{\Large {\bf A Computer Program for Relativistic Multiple Coulomb and
Nuclear Excitation}}

{\Large {\bf \ }}

{\Large \ C.A. Bertulani}

Cyclotron Laboratory, Michigan State Univ., East Lansing, MI 48824-1321, USA%
\footnote{%
Permanent Address, Instituto de F\'\i sica, Universidade Federal do Rio de
Janeiro, 21945-970 Rio de Janeiro, RJ, Brazil. E-mail: bertu@if.ufrj.br}
\end{center}

\begin{quotation}
A computer program is presented by which one may calculate the multiple
electric dipole, electric quadrupole and magnetic dipole Coulomb excitation
with relativistic heavy ions. The program applies to an arbitrary nucleus,
specified by the spins and energies of the levels and by all E1, E2 and M1
matrix elements. Nuclear excitation is calculated optionally for monopole,
dipole and quadrupole excitations and needs inputs of optical potentials.
For given bombarding conditions, the differential cross sections and
statistical tensors (useful to calculate $\gamma$-ray angular distribution
functions) are computed.
\end{quotation}

\section{Introduction}

Relativistic Coulomb Excitation (RCE) is a well established tool to unravel
interesting aspects of nuclear structure \cite{WA79,baur}. The RCE induced
by large-Z projectiles and/or targets, often yields large cross sections in
grazing collisions. This results from the large nuclear response (specially
in the region of the giant resonances) to the acting electromagnetic fields.
As a consequence, a strong coupling between the excited states is expected.
The present report describes a computer program for the calculation of
multiple excitation among a finite number of nuclear states. The system of
coupled differential equations for the time-dependent amplitudes of the
eigenstates of the free nucleus is solved numerically for electric dipole
(E1), electric quadrupole (E2), and magnetic dipole (M1) excitations. If the
optical potential for the system is given, the program can also calculate
the amplitudes for (nuclear) monopole, dipole and quadrupole excitations.
All nuclear quantities, either known from experiments or calculated from a
model, as well as the conditions realized in the experiment, are explicitly
specified as input parameters. The program then computes the Coulomb+nuclear
excitation probabilities and cross sections as well as the statistical
tensors for the angular distribution of the $\gamma $-quanta.

\section{The semiclassical method for the CC-problem}

In relativistic heavy ion collisions, the wavelength associated to the
projectile-target relative motion is much smaller than the characteristic
lengths of the system. It is, therefore, a reasonable approximation to treat 
${\bf r}$ as a classical variable ${\bf r}(t)$, given at each instant by the
trajectory followed by the relative motion. At high energies it is also a
good approximation to replace this trajectory by a straight line. The
intrinsic dynamics can then be handled as a quantum mechanics problem with a
time dependent Hamiltonian. This treatment is discussed in full details by
Alder and Winther in ref.~\cite{AW65}.

The intrinsic state $|\psi (t)>$ satisfies the Schr\"{o}dinger equation 
\begin{equation}
\left\{ H_0\ +\ V\left[ {\bf r}{(t)}\right] \right\} \mid \psi (t)\rangle
=i\hbar {\frac{\partial \mid \psi (t)\rangle }{\partial t}}\;.  \label{eqS}
\end{equation}
Above, $H_0$ is the intrinsic Hamiltonian and $V$ is the channel-coupling
interaction.

Expanding the wave function in the set $\{\mid k\rangle ;\ k=1,N\}$ of
eigenstates of $H_0$, where $N$ is the number states included in the
coupled-channels (CC) problem, we obtain a set of coupled equations. Taking
the scalar product with each of the states $<j|$, we get 
\begin{equation}
i\hbar \ {\dot{a}}_k(t)=\sum_{j=1}^N\ \langle k\mid V(t)\mid j\rangle \;\exp
\left[ i(E_k-E_j)t/\hbar \right] \;a_j(t)\;,\qquad \qquad k,j=1\;{\rm to}%
\;\;N\,.  \label{AW}
\end{equation}
where $E_n$ is the energy of the state $\left| n\right\rangle .$ It should
be remarked that the amplitudes depend also on the impact parameter $b$
specifying the classical trajectory followed by the system. For the sake of
keeping the notation simple, we do not indicate this dependence explicitly.
We write, therefore, $a_n(t)$ instead of $a_n(b,t)$. Since the interaction $%
V $ vanishes as $t\rightarrow \pm \infty $, the amplitudes have as initial
condition $a_n(t\rightarrow -\infty )=\delta _{n1}$ and they tend to
constant values as $t\rightarrow \infty $.

A convenient measure of time is given by the dimensionless quantity $\tau
=\gamma {\rm v}t/b$, where $\gamma =(1-{\rm v}^2/c^2)^{-1/2}$ is the Lorentz
factor for the projectile velocity v. A convenient measure of energy is $%
E_0=\gamma \hbar {\rm v}/b.$ In terms of these quantities the CC equations
become

\begin{equation}
\frac{da_k(\tau )}{d\tau }=-i\sum_{j=1}^N\ \langle k\mid W(\tau )\mid
j\rangle \;\exp \left( i\xi _{kj}\tau \right) \;a_j(\tau )\;;\;\;\;W(\tau )=%
\frac{V(\tau )}{E_0}\;;\;\;\xi _{kj}=\frac{E_k-E_j}{E_0}\;\;.  \label{AW2}
\end{equation}

The nuclear states are specified by the spin quantum numbers $I$ and $M$.
Therefore, the excitation probability of an intrinsic state $\mid n\rangle
\equiv \mid I_n,M_n\rangle $ in a collision with impact parameter $b$ is
obtained from an average over the initial orientation $\left( M_1\right) ,$
and a sum over the final orientation of the nucleus, respectively: 
\begin{equation}
P_n(b)=\frac 1{2I_1+1}\sum_{M_1,M_n}|a_{I_n,M_n}(M_1)|^2\;.  \label{Pn}
\end{equation}
The total cross section for excitation of the state $|n>$ is obtained by the
classical expression 
\begin{equation}
\sigma _n=2\pi \ \int \ P_n(b)\ b\;db\;.
\end{equation}

\subsection{Coulomb excitation}

We consider a nucleus 2 which is at rest and a relativistic nucleus 1 which
moves along the $z$-axis. Nucleus 2 is excited from the initial state $%
|I_jM_j>$ to the state $|I_kM_k>$ by the electromagnetic field of nucleus 1.
The nuclear states are specified by the spin quantum numbers $I_j$, $I_k$
and by the corresponding magnetic quantum numbers $M_j$ and $M_k$. We assume
that the relativistic nucleus 1 moves along a straight-line trajectory with
impact parameter $b$, which is therefore also the distance of the closest
approach between the center of mass of the two nuclei at the time $t=0$. The
interaction, $V_C(t),$ due to the electromagnetic field of the nucleus 1
acting on the charges and currents nucleus 2 can be expanded into
multipoles, as explained in ref. \cite{BC96}. One has 
\begin{equation}
W_C(\tau )=\frac{V_C(\tau )}{E_0}=\sum_{\pi \lambda \mu }W_{\pi \lambda \mu
}(\tau )\ ,  \label{Vfi3}
\end{equation}
where $\pi =E,\;M$ denotes electric and magnetic interactions, respectively,
and (misprints in ref. \cite{BC96} have been corrected)

\begin{equation}
W_{\pi \lambda \mu }(\tau )=\left( -1\right) ^{\lambda +1}\frac{Z_1e}{\hbar 
{\rm v}b^\lambda }\frac 1\lambda \;\sqrt{\frac{2\pi }{\left( 2\lambda
+1\right) !!}}\;Q_{\pi \lambda \mu }(\xi ,\tau )\;{\cal M}(\pi \lambda ,\mu
)\;,  \label{wlamb}
\end{equation}
where ${\cal M}(\pi \lambda ,\mu )$ is the multipole moment of order $%
\lambda \mu ,$

\begin{equation}
{\cal M}(E\lambda ,\mu )=\int d^3r\ \rho ({\bf r})\ r^\lambda \ Y_{1\mu }(%
{\bf r})\ ,  \label{ME1}
\end{equation}
and

\begin{equation}
{\cal M}(M1,\mu )=-{\frac i{2c}}\ \int d^3r\ {\bf J}({\bf r}).{\bf L}\left(
rY_{1\mu }\right) \ ,  \label{MM1}
\end{equation}
$\rho $ (${\bf J}$) being the nuclear charge (current). The quantities $%
Q_{\pi \lambda \mu }(\tau )$ were calculated in ref. \cite{BC96}, and for
the E1, E2, and M1 multipolarities are given by

\begin{equation}
Q_{E10}(\xi ,\tau )=\gamma \sqrt{2}\left[ \tau \phi ^3(\tau )-i\xi \left( 
\frac{{\rm v}}c\right) ^2\phi (\tau )\right] \;;\;\;\;\;\;Q_{E1\pm 1}(\xi
,\tau )=\mp \phi ^3(\tau )\;,  \label{QE1}
\end{equation}
\begin{equation}
Q_{M10}(\xi ,\tau )=0\;;\;\;\;\;\;Q_{M1\pm 1}(\xi ,\tau )=i\left( \frac{{\rm %
v}}c\right) \phi ^3(\tau )\;,  \label{QM1}
\end{equation}
and

\begin{eqnarray}
Q_{E20}(\xi ,\tau ) &=&\gamma ^2\sqrt{6}\left[ \left( 2\tau ^2-1\right) \phi
^5(\tau )-i\xi \left( \frac{{\rm v}}c\right) ^2\tau \phi ^3\left( \tau
\right) \right] \;;\;\;\;  \nonumber  \label{QE2} \\
\;\;Q_{E2\pm 1}(\xi ,\tau ) &=&\pm \gamma \left[ 6\tau \phi ^5(\tau )-i\xi
\left( \frac{{\rm v}}c\right) ^2\phi ^3\left( \tau \right) \right]
\;;\;\;\;\;Q_{E2\pm 2}(\tau )=3\phi ^5(\tau )\;,  \label{QE20}
\end{eqnarray}
where $\phi \left( \tau \right) =\left( 1+\tau ^2\right) ^{-1/2}.$

We use here the notation of Edmonds \cite{Ed60} where the reduced multipole
matrix element is defined by

\begin{equation}
{\cal M}_{kj}(E\lambda ,\mu )=(-1)^{I_k-M_k}\ \left( {{{{{I_k \atop -M_k}}}%
}}{{{{{\lambda \atop \mu }}}}}{{{{{I_j\atop M_j}}}}}\right) <I_k||{\cal M%
}(E\lambda )||I_j>\   \label{Mfi1}
\end{equation}
To simplify the expression (\ref{AW2}) we introduce the dimensionless
parameter $\psi _{kj}^{(\lambda )}$ by the relation

\[
\psi _{kj}^{(\lambda )}=\left( -1\right) ^{\lambda +1}\frac{Z_1e}{\hbar {\rm %
v}b^\lambda }\frac 1\lambda \sqrt{\frac{2\pi }{\left( 2\lambda +1\right) !!}}%
\;{\cal M}_{kj}(E\lambda ) 
\]
Then we may write eq. \ref{AW2} in the form

\begin{equation}
\frac{da_k(\tau )}{d\tau }=-i\sum_{r=1}^N\ \sum_{\pi \lambda \mu }Q_{\pi
\lambda \mu }(\xi _{kj},\tau )\psi _{kj}^{\left( \lambda \right) }\;\exp
\left( i\xi _{kj}\tau \right) \;a_j(\tau )\;.  \label{ats1}
\end{equation}

The fields $Q_{\pi \lambda \mu }(\xi ,\tau )$ peak around $\tau =0$, and
decrease rapidly within an interval $\Delta \tau \simeq 1$, corresponding to
a collisional time $\Delta t\simeq b/\gamma {\rm v}$. This means that
numerically one needs to integrate the CC equations in time within an
interval of range $n\times \Delta \tau $ around $\tau =0$, with $n$ equal to
a small integer number.

\subsection{Nuclear excitation}

In peripheral collisions the nuclear interaction between the ions can also
induce excitations. According to the Bohr-Mottelson particle-vibrator
coupling model, the matrix element for the transition $j\longrightarrow k$
is given by 
\begin{equation}
V_{N(\lambda \mu )}^{(kj)}({\bf r})\equiv <I_kM_k|V_{N(\lambda \mu
)}|I_jM_j>={\frac{\delta _\lambda }{\sqrt{2\lambda +1}}}\ <I_kM_k|Y_{\lambda
\mu }|I_jM_j>\ Y_{\lambda \mu }(\hat{{\bf r}})\ U_\lambda (r)  \label{VfiN}
\end{equation}
where $\delta _\lambda $is the vibrational amplitude and $U_\lambda (r)$ is
the transition potential.

The transition potentials for nuclear excitations can be related to the
optical potential in the elastic channel. This is discussed in details in
ref. \cite{Sa87}. The transition potentials for isoscalar excitations are 
\begin{equation}
U_0(r)=3U_{opt}(r)+r{\frac{dU_{opt}(r)}{dr}}\ ,  \label{U0}
\end{equation}
for monopole, 
\begin{equation}
U_1(r)={\frac{dU_{opt}}{dr}}+{\frac 13}\ R_0\ {\frac{d^2U_{opt}}{dr^2}}\ ,
\label{U1}
\end{equation}
for dipole, and

\begin{equation}
U_2(r)={\frac{dU_{opt}(r)}{dr}}\ ,  \label{U2}
\end{equation}
for quadrupole modes.

The time dependence of the matrix elements above can be obtained by making a
Lorentz boost. One gets 
\begin{eqnarray}
V_{N(\lambda \mu )}^{(kj)}(t) &\equiv &<I_kM_k|U|I_jM_j>  \nonumber \\
&=&\gamma \ {\frac{\delta _\lambda }{\sqrt{2\lambda +1}}}\
<I_kM_k|Y_{\lambda \mu }|I_jM_j>Y_{\lambda \mu }\left( \theta (t){,0}\right)
\ U_\lambda [r(t)]\ ,  \label{VfiN2}
\end{eqnarray}
where $r(t)=\sqrt{b^2+\gamma ^2{\rm v}^2t^2}=1/\left( b\phi \left( \tau
\right) \right) ,$ $\theta =\tau \phi (\tau )$, and

\begin{equation}
<I_kM_k|Y_{\lambda \mu }|I_jM_j>=(-1)^{I_k-M_k}\ \left[ {\frac{%
(2I_k+1)(2\lambda +1)}{4\pi (2I_j+1)}}\right] ^{1/2}\ \left( {{{{{I_k\atop %
-M_k}}}}}{{{{{\lambda\atop \mu }}}}}{}{{{{{I_j\atop M_j}}}}}\right)
\left( {{{{{I_k\atop 0}}}}}{{{{{\lambda \atop 0}}}}}{{{{{I_j \atop 0}}}%
}}\right) \ .  \label{WE}
\end{equation}

To put it in the same notation as in eq. (\ref{ats1}), we define $%
Q_{N\lambda \mu }^{(kj)}(\tau )=V_{N(\lambda \mu )}^{(kj)}(t)/E_0$, and the
coupled-channels equations become

\begin{equation}
\frac{da_k(\tau )}{d\tau }=-i\sum_{j=1}^N\ \sum_{\lambda \mu }\left[
Q_{N\lambda \mu }^{(kj)}(\tau )+\sum_\pi Q_{\pi \lambda \mu }(\xi _{kj},\tau
)\psi _{kj}^{\left( \lambda \right) }\right] \;\exp \left( i\xi _{kj}\tau
\right) \;a_j(\tau )\;.  \label{ats2}
\end{equation}

\subsection{Absorption at small impact parameters}

If the optical potential $U_{opt}({\bf r})$ is known, the absorption
probability in grazing collisions can be calculated in the eikonal
approximation as

\begin{equation}
A(b)=\exp \left[ \frac 2{\hbar {\rm v}}\int_{-\infty }^\infty {\rm Im}\left[
U_{opt}({\bf r})\right] dz\right] \;,  \label{abs}
\end{equation}
where $r=\sqrt{b^2+z^2}$. If the optical potential is not known, the
absorption probability can be calculated from the optical limit of the
Glauber theory of multiple scattering, which yields:

\begin{equation}
A(b)=\exp \left\{ -\sigma _{NN}\int_{-\infty }^\infty \ \left[ \int \rho _1(%
{\bf r}^{\prime })\ \rho _2({\bf r-r^{\prime }})\ d^3r^{\prime }\right]
dz\right\} \;.  \label{abs2}
\end{equation}
where $\sigma _{NN}=40$ mb is the nucleon-nucleon cross section in high
energy collisions and $\rho _i$ is the ground state density of the nucleus $%
i.$ These densities are taken from the droplet model densities of Myers and
Swiatecki \cite{MS69}, but can be easily replaced by more realistic
densities.

Including absorption, the total cross section for excitation of the state $%
|n>$ is obtained by 
\begin{equation}
\sigma _n=2\pi \ \int \ A(b)P_n(b)\ bdb\;.  \label{sigman2}
\end{equation}

\subsection{Coulomb recoil correction}

At intermediate bombarding energies a correction of the excitation
amplitudes due to Coulomb recoil can be easily introduced. The excitation
occurs most probably when the nuclei are closer along the trajectory. But
then, due to recoil, they are displaced by an extra distance of order of $%
a_0/\gamma $, where $a_0=Z_1Z_2e^2/m_0{\rm v}^2$. It is thus expected that
one should correct the amplitudes by a rescaling of the impact parameter. In
ref. \cite{WA79} it was shown that a comparison between the Rutherford
integrals with large angular momenta for the non-relativistic case with the
straight-line integrals for the relativistic case yields a reasonable
correction. It amounts in the replacement of $b$ of $P_n(b)$ in equation (%
\ref{sigman2}) (and also in $A(b)$ of eq. (\ref{abs}))
by $b^{\prime }=b+\pi a_0/2\gamma $, i.e., for a given impact
parameter $b$ the excitation amplitudes and probabilities are calculated for 
$b^{\prime }.$

\subsection{Gamma-ray angular distributions}

As for the non-relativistic case \cite{WB65,AW65}, the angular distributions
of gamma rays following the excitation depend on the frame of reference
used. In our notation, the z-axis corresponds to the beam axis, and the
statistical tensors are given by

\begin{eqnarray}
A_{k\kappa }(N) &=&\frac{\left( 2I_N+1\right) ^{1/2}}{\left( 2I_1+1\right) }%
\sum_{M_N=-(M_N^{^{\prime }}+\kappa),M_N^{^{\prime }}}\left( -1\right)
^{I_N+M_N}\left( 
\begin{tabular}{lll}
$I$$_N$ & $I$$_N$ & $k$ \\ 
$M$$_N$ & $M$$_N^{^{\prime }}$ & $\kappa $%
\end{tabular}
\right)  \nonumber \\
&\times &\sum_{M_1}a_{I_NM_N^{^{\prime }}}^{*}(M_1)\;a_{I_NM_N}(M_1)\;,
\label{stat}
\end{eqnarray}
where $N$ is the state from which the gamma ray is emitted, and $0$ denotes
the initial state of the nucleus, before the excitation. To calculate the
angular distributions of the gamma rays one needs the statistical tensors
for $k=0,2,4$ and $-k\leq \kappa \leq k.$ Since the nuclear levels are not
only populated by Coulomb excitation but also by internal conversion and
gamma transitions cascading down from higher states, the angular
distributions depend not only on the $\gamma $-transition rates, $\delta
_{N\rightarrow M}$ (for all $M$'s), but also on the internal conversion
coefficients, $\alpha _{N\rightarrow M}$. With those parameters at hand one
calculates the angular distributions of gamma rays, $dW_{\gamma N\rightarrow
M}/d\Omega _\gamma $, by using the equations given in section IV of ref. 
\cite{WB65} (also reprinted in \cite{AW65}).

\section{Computer program and user's manual}

The units used in the program are fm (Fermis) for distances and MeV for
energies. The output cross sections are given in millibarns.

\subsection{Input parameters}

To avoid exceeding use of computer's memory, the file RELEX.DIM contains the
dimension of the arrays and sets in the maximum number of levels (NMAX),
maximum total number of magnetic substates, (NSTMAX), maximum number of
impact parameters (NBMAX), and maximum number of coordinates points used in
the optical potentials and absorption factors, (NGRID).

The integrals appearing in eqs. (22-24) are performed by the 1/3-Simpson's
integration rule. It is required that NGRID be a {\it even} number, since an
extra point (origin) is generated in the program.

The file RELEX.IN contains all other input parameters. These are

\begin{enumerate}
\item  AP, ZP, AT, ZT, which are the projectile and the target mass and
charge numbers, respectively.

\item  ECA, the bombarding energy per nucleon in MeV.

\item  EX(j) and SPIN(j): the energy and spins of the individual states j.

\item  MATE1(j,k), MATE2(j,k), MATM1(j,k), the reduced matrix elements for
E1, E2 and M1 excitations, $j\rightarrow k,$ \ (as defined in (\ref{Mfi1})).

\item  DELTE(0,j), DELTE(1,j), DELTE(2,j), the deformation parameters for
monopole, dipole and quadrupole nuclear excitations, entering eq. (\ref
{VfiN2}).
\end{enumerate}

The input cards in file RELEX.IN are organized as following:

\begin{enumerate}
\item  First data card: AP, ZP, AT, ZT, ECA, IW, IOUT.

IW = 0 (or 1) for projectile (or target) excitation

IOUT = 0 (or 1) for output (or not) of statistical tensors. The statistical
tensors are calculated for each impact parameter, so that one can use them
in the computation of $dP_{\gamma N\rightarrow M}(b)/d\Omega _\gamma
=P_N(b).dW_{\gamma N\rightarrow M}/d\Omega _\gamma $.

\item  Second data card: NB, ACCUR, BMIN, ITOT

NB = number of points in impact parameter mesh.

ACCUR = accuracy required for the time integration of the CC-equations for
each impact parameter. A reasonable value is ACCUR = 0.001, i.e., 0.1\%.

BMIN = minimum impact parameter (can be set to zero).

ITOT = 1 (0) prints (does not print) out impact parameter probabilities.

\item  Third data card: IOPW, IOPNUC

IOPW is an option to use (or not) an optical potential: IOPW = 1 (0).

If the optical potential is provided (IOPW = 1), it should be stored in file
IOPW.IN in rows of R x Real[U(R)] x Imag[U(R)]. The first row in this file
gives the number of rows (maximum = NGRID). The program makes an
interpolation to obtain intermediate values.

IOPNUC = 1 (0) is an option to compute (or not) nuclear excitation.

\item  Fourth data card: NST

Number of nuclear levels.

\item  Fifth and following data cards: J, EX(J), SPIN(J)

Input of state labels (J), energy (EX), and angular momentum (SPIN). J
ranges from 1 to NST and should be listed in increasing value of energies.

\item  Following data cards: J, K, MATE1(J,K), MATE2(J,K), MATM1(J,K)

Reduced matrix elements for electromagnetic E1, E2 and M1 excitations:

MATE1(j$\rightarrow $k), for electric dipole transitions, MATE2(j$%
\rightarrow $k), for electric quadrupole transitions, MATM1(j$\rightarrow $%
k), for magnetic dipole transitions. Matrices for reorientation effects, i$%
\rightarrow $i, can also be given.

To end reading the matrix elements add a row of zeros for each column.

\item  Following data cards: K, DELTE (0,K), DELTE(1,K), DELTE(2,K)

If IOPNUC = 1 these rows give the values of deformation parameters for
monopole, dipole and quadrupole nuclear excitations, respectively, for each
excited state K.
\end{enumerate}

\subsection{Computer program}

The program starts with a catalogue of the nuclear levels by doing a
correspondence of integers to each magnetic substate. J = 1 corresponds to
the lowest energy level, with the magnetic quantum number $M_1=-I_1$. J
increases with $M_1$ and so on for the subsequent levels.

A mesh in impact parameter is done, reserving half of the impact parameter
points, i.e., $NB/2$, to a finer mesh around the grazing impact parameter,
defined as $b_0=1.2\left( A_P^{1/3}+A_T^{1/3}\right) \;$fm$.$ The interval $%
b_0/2\;fm\leq b\leq 3b_0/2\;fm$ is covered by this mesh. A second mesh, with
the other half of points, extends from $b=3b_0/2\;fm$ to $b=200\;fm.$ Except
for the very low excitation energies ($E_j\ll 1$ MeV), combined with very
large bombarding energies ($\gamma \gg 1$), this upper value of $b$
corresponds to very small excitation probabilities, and the calculation can
be safely stopped. The reason for a finer mesh at small impact parameters is
to get a good accuracy at the region where both nuclear, Coulomb, and
absorption factors play equally important roles. At large impact parameters
the probabilities fall off smoothly with $b$, justifying a wider integration
step.

A mesh of NGRID points in polar coordinates is implemented to calculate the
nuclear excitation potentials and absorption factors, according to the
equations presented in sections 2.2 and 2.3. The first and second
derivatives of the optical potentials are calculated by the routine
DERIVATIVE. A 6-point formula is used for the purpose. The routines TWOFOLD
computes the folding over the densities, as used in eq. (\ref{abs2}).
Routines RHOPP and RHONP generate the liquid drop densities, and the routine
PHNUC computes the eikonal integral appearing in eq. (\ref{abs}).

Repeated factors for the nuclear and for the Coulomb potentials are
calculated in the main program and stored in the main program with the
arrays PSITOT and PSINUC. These arrays are carried over in a common block to
the routine VINT which computes the function $Q_{N\lambda \mu }^{(kj)}(\tau
)+\sum_\pi Q_{\pi \lambda \mu }(\xi _{kj},\tau )\psi _{kj}^{\left( \lambda
\right) }$, used in eq. (\ref{ats2}).

The time integrals are performed by means of an adaptive Runge-Kutta method.
All routines used for this purpose have been taken from the Numerical
Recipes Software, described in the book \cite{Numrep}. They are composed by
the routines ODEINT, RKQS, RKCK, and RK4. The routine ODEINT varies the time
step sizes to achieve the desired accuracy, controlled by the input
parameter ACCUR. The right side of (\ref{ats2}) is computed in the routine
DCADT, used externally by the fourth-order Runge-Kutta routine RK4. RKCK is
a driver to increase time steps in RK4, and RKQS is used in ODEINT for the
variation of step size and accuracy control. The main program returns a
warning if the summed errors for all magnetic substates is larger than 10 $%
\times $ ACCUR.

The routine THREEJ computes Wigner-3J coefficients (and Clebsh-Gordan
coefficients), and YLM is used to compute the spherical harmonics$.$

The routines SPLINE and SPLINT perform a spline interpolation of the
excitation amplitudes, before they are used for integration by means of the
routines QTRAP and QSIMP. All of them are from Numerical Recipes Software 
\cite{Numrep}.

The program delivers outputs in RELEX.OUT. The statistical tensors are given
in RELEX2.OUT for each impact parameter (only if IOUT=1).

The program is available at \underline{%
http://www.if.ufrj.br/~bertu/prog.html} .

\bigskip\bigskip

{\bf Acknowledgments}

I would like to express my gratitude to Heiko Scheit, Gregers Hansen and
Thomas Glasmacher for useful comments and suggestions during the development
of this work. This work was partially supported by the Brazilian agencies:
CNPq, FAPERJ, FUJB, PRONEX, and by the Michigan State University.

\end{document}